\documentclass[sigconf,review=false, screen, anonymous=false,usenames,dvipsnames,svgnames,table]{acmart}

\usepackage{graphicx}
\usepackage{caption}
\usepackage{pbox}
\usepackage{url}
\usepackage{hyperref}
\usepackage{multirow}
\usepackage{natbib}
\usepackage{subcaption}

\usepackage{tabularx}

\usepackage{todonotes}

\usepackage[strict]{changepage}

\usepackage{framed}

\definecolor{formalshade}{rgb}{0.95,0.95,1}

\makeatletter
\newcommand{\interviewquote}[2]{
 \def\FrameCommand{%
    \hspace{0pt}%
    {\color{MidnightBlue}\vrule width 1.5pt}%
    {\color{white}\vrule width 4pt}%
    \colorbox{white}
  }%
  \MakeFramed{\advance\hsize-\width\FrameRestore}%
  \noindent\hspace{-4.55pt}
  \begin{adjustwidth}{}{7pt}
  \footnotesize{"\emph{#1}" -{#2}}\vspace{0.5pt}\end{adjustwidth}\endMakeFramed%
}
\makeatother

\definecolor{javagreen}{rgb}{0.25,0.5,0.35} 

\definecolor{purple}{rgb}{0.5,0,0.5} 

\definecolor{blue}{rgb}{0,0.7,0.9} 

\ccsdesc[300]{Software and its engineering~Software notations and tools}

\begin{document}

\settopmatter{printacmref=true}

\title{An Empirical Study of Bots in Software Development: Characteristics and Challenges from a Practitioner's Perspective}

\author{Linda Erlenhov}
\affiliation{%
  \institution{Software Engineering Division,\\Chalmers $|$ University of Gothenburg}
  \city{Gothenburg}
  \country{Sweden}
}
\email{linda.erlenhov@chalmers.se}
\author{Francisco Gomes de Oliveira Neto}
\affiliation{%
  \institution{Software Engineering Division,\\Chalmers $|$ University of Gothenburg}
  \city{Gothenburg}
  \country{Sweden}
}
\email{francisco.gomes@cse.gu.se}
\author{Philipp Leitner}
\affiliation{%
  \institution{Software Engineering Division,\\Chalmers $|$ University of Gothenburg}
  \city{Gothenburg}
  \country{Sweden}
}
\email{philipp.leitner@chalmers.se}

\copyrightyear{2020}
\acmYear{2020}
\setcopyright{acmlicensed}\acmConference[ESEC/FSE '20]{Proceedings of the 28th ACM Joint European Software Engineering Conference and Symposium on the Foundations of Software Engineering}{November 8--13, 2020}{Virtual Event, USA}
\acmBooktitle{Proceedings of the 28th ACM Joint European Software Engineering Conference and Symposium on the Foundations of Software Engineering (ESEC/FSE '20), November 8--13, 2020, Virtual Event, USA}
\acmPrice{15.00}
\acmDOI{10.1145/3368089.3409680}
\acmISBN{978-1-4503-7043-1/20/11}

%
%
%
%
%
 \keywords{Software bot, Empirical study, Software engineering}


\begin{abstract}
Software engineering bots -- automated tools that handle tedious tasks -- are increasingly used by industrial and open source projects to improve developer productivity. Current research in this area is held back by a lack of consensus of what software engineering bots (DevBots) actually are, what characteristics distinguish them from other tools, and what benefits and challenges are associated with DevBot usage. In this paper we report on a mixed-method empirical study of DevBot usage in industrial practice. We report on findings from interviewing 21 and surveying a total of 111 developers. We identify three different personas among DevBot users (focusing on autonomy, chat interfaces, and ``smartness''), each with different definitions of what a DevBot is, why developers use them, and what they struggle with. We conclude that future DevBot research should situate their work within our framework, to clearly identify what type of bot the work targets, and what advantages practitioners can expect. Further, we find that there currently is a lack of general-purpose ``smart'' bots that go beyond simple automation tools or chat interfaces. This is problematic, as we have seen that such bots, if available, can have a transformative effect on the projects that use them.
\end{abstract}

\maketitle


\section{Introduction}
\label{sec:intro}

Modern software development entails a wide range of tasks that go beyond designing and implementing program code. Among many other tasks, developers may be asked to conduct code reviews~\cite{bacchelli:13}, triage bugs~\cite{baysal:12}, mentor newcomers~\cite{canfora:12}, run integration or performance tests and interpret test result~\cite{costa:19}, set up and maintain computational infrastructure~\cite{zhu:16}, or stay ``on call'' to react to production issues~\cite{feitelson:13}. Unsurprisingly, previous work has reported that developers often feel distracted and unproductive, especially if they have to change context frequently~\cite{meyer14}.

Partly as a reaction, more and more software projects employ software development bots (DevBots), automated tools that attempt to free developers from particularly tedious tasks, or support their work in a more general sense.
DevBots are by now wide-spread in industry and open source software (e.g., Wessel et al. found no less than 48 different bots being used in 93 open source projects~\cite{wessel:18}), and research on bots in software engineering is gaining traction (see for instance the BotSE workshop at the ICSE conference\footnote{\url{http://botse.org}}).

Unfortunately, research so far has been held back by a lack of consensus of what a bot in software engineering actually is, which characteristics define DevBots, and why developers use them. Early studies have led to a multitude of taxonomies and classification attempts~\cite{lebeuf2018ieee,erlenhov2019botse,paikari:18}. While useful, these are largely based on a categorisation of existing tools that their authors refer to as ``bots'', without explicitly incorporating the view of bot users.

In contrast, in this paper, we describe results of a study investigating the perception of DevBots from the side of the practitioners using them. We carried out a mixed-method exploratory study over a period of 6 months. The goal of our study was to identify which characteristics distinguish DevBots from standard tools (subsequently called ``Plain Old Development Tools'', or PODTs), and to improve the community's understanding of DevBot usage and challenges in industrial practice.
We conducted semi-structured interviews with 21 software developers, who utilise a wide range of different DevBots in their work, and enhance our data through a Web-based survey answered by 111 professional developers or other IT professionals, 60 of which are or have been using DevBots in their work. Concretely, we address the following research questions:

\vspace{0.25cm}
\emph{\textbf{RQ1}: What is a DevBot? What characteristics describe a DevBot? What delineates DevBots from Plain Old Development Tools (PODTs)?}\\

Our study shows that a single definition of DevBots is unachievable, as different developers associate widely different characteristics with the term. However, we are able to identify three different personas, i.e., practitioner archetypes with different expectations and motivations~\cite{pruitt:03}. The chat bot persona (Charlie) sees DevBots mostly as (information) integration tools with a natural language interface, while for the autonomous bot persona (Alex) a DevBot is a tool that autonomously handles, often quite simple, tasks for human developers. Finally, for the smart bot persona (Sam), the distinguishing characteristic of a bot is a ``smartness'' that goes beyond other tools. By classifying survey response data, we are able to estimate the prevalence of these personas in the wild. We find that 48\% of respondents who are actually using DevBots in their work are predominantly associated with Charlie, followed by Alex (19\%) and Sam (13\%). The remaining 20\% of bot-using respondents gave responses that did not allow us to map them to any of these three personas.

\vspace{0.25cm}

\emph{\textbf{RQ2}: How are DevBots used? What benefits do different types of DevBots promise?}\\

All DevBot users primarily expect improved productivity from their bots, but depending on the persona this surfaces in different ways. Charlie mainly expects easy access to information or being able to trigger simple maintenance tasks through a natural language interface. Conversely, Alex expects that simple-but-tedious tasks are automated, \emph{without} explicitly having to trigger the bot. Finally, Sam also expects improved productivity, but in a less straight-forward manner --- for Sam, improved productivity comes from DevBots handling non-trivial tasks or generating information that would otherwise not easily be accessible to a human.

\vspace{0.25cm}

\emph{\textbf{RQ3}: What are the main challenges in DevBot usage?}\\

DevBots as defined by Charlie are widely available in practice, but their usage is sometimes subject to usability concerns: bots that can parse rich natural language are perceived as unpredictable, while simple bots that only ``understand'' a small set of defined trigger words or sentences are seen as less useful. Alex and Sam need to be able to trust their bots to autonomously trigger and enact correct actions. This requires mature bot implementations with a very small amount of false positives, as well as trustworthy test suites which are able to alert developers quickly in case of malfunctioning bots. Charlie and Alex-style bots are already widely available as off-the-shelf systems. Sam-style bots, on the other hand, are bespoke tools for individual projects or companies. In our study we have only observed their usage in large software-intensive corporations, which invest considerable effort into building and maintaining tailor-made bots for their own environments.

\vspace{0.25cm}

Our results provide clear terminology and definitions to better contextualise and situate future research on bots in software engineering. Further, our results indicate that there is currently a lack of general-purpose Sam-style DevBots. This is problematic, as adopting more sophisticated DevBots can have transformative effects that go beyond simply automating tedious work items.


\section{Related Work}
In 2016, Storey and Zagalsky laid the foundation for research on bots in software engineering. They described how bots are increasingly used to support tasks that traditionally required human intelligence~\cite{storey:16}. This early work already established that DevBots may come in very different forms, support a wide range of use cases, and occupy different roles in software teams. Lebeuf and Storey later explored this variety further through an extensive faceted taxonomy of bots~\cite{lebeuf2018ieee,lebeuf2019}, which illustrated how completely different tools may all be perceived as ``bots'' by different developers. An alternative taxonomy proposed later by Erlenhov et al.~\cite{erlenhov2019botse} similarly indicates that different qualities may be characteristic of bots to different people.
 A third taxonomy has been proposed by Paikari and van der Hoek~\cite{paikari:18} and specifically investigates chat bots in software engineering and beyond. In contrast to these taxonomies, our study tackles a related, yet not identical, question: what exactly characterises bots, and in what aspects they are perceived as different from PODTs. Further, and unlike the taxonomies proposed by Lebeuf and Storey or Paikari and van der Hoek, we focus solely on bots used for software engineering tasks. Finally, we conduct our research from the perspective of bot users, rather than bot developers and vendors.

Wessel et al. have shown that DevBot usage is indeed widespread, at least in the context of open source software development~\cite{wessel:18}. Through repository mining they found 48 different bots being used in 93 open source projects. Over one fourth of all analyzed projects used at least one bot. However, they were not able to show a clear positive impact of these bots on high-level project quality metrics, such as the number of commits or pull request merge times. We speculate that one problem may be that we simply do not yet understand well enough why projects adopt specific bots -- potentially many of these bots  were never intended to improve the metrics that Wessel et al. studied. Our study attempts to shine more light on what benefits developers expect from different types of DevBots.

Other research has indicated further potential challenges of using DevBots. For instance, a well-known experiment by Murgia et al. indicated that users are sceptical of bot contributions on the developer Q\&A site Stack Exchange~\cite{murgia2016}. Recently, this has also been confirmed by Brown and Parnin through an experiment with bot-generated contributions to open source projects~\cite{brown:19} --- from 52 pull requests submitted by a DevBot, only two were merged in their experiment (and these two were later reverted by the project owners). These experiences have led some bot developers to camouflage their bots as human developers. For instance, the program repair bot Repairnator~\cite{urli:18} has, for a while, submitted pull requests using a human profile to improve acceptance. While understandable, this practice may potentially be harmful to software projects. Ferrara et al. have discussed the threat of malicious, non-obvious bots damaging online ecosystems~\cite{ferrara:16}. In other large online ecosystems with longer experience using bots (e.g., Wikipedia), rigid governance rules have been established~\cite{mueller:13}. Wikipedia bots need to contain the string ``bot'' in their name, have a discussion page that clearly describes what they do, and can be turned off by any member of the community at any time. In our research, we investigate what problems developers face when using bots. We explicitly study under which conditions developers trust bot-generated contributions.

Finally, various projects have proposed DevBots for a wide range of software engineering tasks. Examples include agile team management~\cite{ablett:07,matthies:19}, program repair~\cite{urli:18,vantonder:19}, software visualization~\cite{bieliauskas:17}, source code refactoring~\cite{wyrich:19}, or pull request management~\cite{wessel:19}. Our work is orthogonal to these studies, as we are not proposing any concrete new type of DevBot. Instead, the subject of our work is how developers define and perceive DevBots in general.


\section{Study Methodology}
\label{sec:method}

%

To address our research questions, we conduct a study based on techniques found in Grounded Theory \cite{stol2016icse,hoda12} (GT), namely coding, memoing, sorting, constant comparison and theoretical saturation.
We cannot claim to use a complete GT method because we had wide exposure to literature prior to beginning our analysis, such that some of our themes align with facets of existing taxonomies for bots in software development~\cite{stol2016icse}.

In order to strengthen our findings, we follow the recommendations by Bratthall and Jorgensen~\cite{bratthall02} and use a methodology that consists of two isolated steps of data collection and iterative phases of data analysis.
First, we defined a set of open-ended questions from our research questions and conducted \textit{qualitative, semi-structured interviews} with 21 participants.
Second, to further substantiate our findings, we ran a \textit{quantitative, Web -based survey} and gathered responses from 111 professional software developers and other IT professionals.
Interview guide, survey materials, and analysis scripts can be found in the replication package of the study~\cite{replicationPackage}. The replication package does not contain interview transcripts to preserve participant privacy.

\subsection{Interviews}

We used our research questions to structure our interview guide.
When conducting the interviews, we followed the flow of the interview, and
gave our participants some freedom when describing how they perceive and use DevBots.

\subsubsection*{Participants:} We sampled industry practitioners that, at some point, worked with tools that they self-identified as DevBots. We began by inviting practitioners from our personal industry network, who then referred us further to other potential participants. Then, we used a saturation approach~\cite{menzies2016book} where we kept inviting new participants in parallel to data analysis while the data offered new information. An overview of the participants is found in Table~\ref{tab:participants}. Most of our participants reported working in corporations of 250 employees or more, but even within this group there is a spectrum of companies spanning organisations between 250 to 10000+ employees. Further, our interviewees span various domains and professional roles, and reported on average 11 years ($\pm 5$) of professional experience.

\begin{table}[h!]
  \scriptsize
  \centering
  \caption{Interview Study Participants}
  \vspace{-0.3cm}
  \label{tab:participants}
  \begin{tabular}{cllc}
    \toprule
    ID  &  Role & Company Size & Experience (Years)\\
    \midrule
    I1 & Software Tester & Corporation & 1.5 \\
    I2 & Senior Software Developer & Corporation & 15 \\
    I3 & Data Scientist & Corporation & 9\\
    I4 & Senior Developer & Corporation & 16\\
    I5 & Senior Developer & Corporation & 15\\
    I6 & Security Architect & Corporation & 20\\
    I7 & Software Engineer & Corporation & 15\\
    I8 & DevOps Engineer & Corporation & 12\\
    I9 & Lead Data Engineer & Corporation & 12\\
    I10 & Product Owner & Corporation & 8\\
    I11 & CEO / Founder & Startup & 7\\
    I12 & Software Developer & Single person & 15\\
    I13 & Software Engineer & SME & 3.5\\
    I14 & Software Quality Consultant & SME & 6\\
    I15 & VP of Engineering & SME & 4\\
    I16 & Software Developer & Corporation & 15\\
    I17 & Data Science Consultant & Corporation & 17\\
    I18 & Devops Engineer & Corporation & 3\\
    I19 & Software Engineer & Corporation & 15\\
    I20 & Test Automation Engineer & Corporation & 6\\
    I21 & Software Engineering Manager & Corporation & 20\\
  \bottomrule
\end{tabular}
\end{table}

\subsubsection*{Protocol:} We conducted interviews over a period of three months. Each interview took between 15 and 45 minutes and was done either face-to-face or via video conference.
Prior to each interview, participants were asked for consent to record the interview and use their data in our study. Participants were offered the opportunity to opt out of the study at any point. However, none of the participants dropped from the study.

\subsubsection*{Analysis:} In parallel to data collection, we performed open and axial coding based on the Straussian variant of GT~\cite{stol2016icse,strauss1998basics}.
In open coding we fracture the data to find relevant excerpts.
In axial coding, we aggregate and connect those excerpts into categories and themes until achieving saturation.
This also allows us to identify limitations of the interview guide, such as missing relevant aspects that were not clear upon definition of the research question.
For instance, during our first coding round, we identified that many participants discussed the issue of trust in DevBots, even though we did not include an explicit question on the topic.
Consequently, we were able to update the interview guide to include one additional question about challenges with trust in DevBots.

In order to triangulate results, different pairs of authors performed open coding on the first 4 interviews to check for consistency and agreement in our coding process. A total of 13 interviews were openly coded independently by two authors, whereas the remaining 8 interviews were open coded by only one author. The second part, the axial coding in which we identified themes, memoed and performed card sorting, was done by all authors together in different sessions lasting between 2--3 hours each. The resulting categories and findings are supported by statements from multiple participants.

\subsection{Survey}

In the second step of our study, we designed a Web-based survey using Typeform\footnote{\url{https://www.typeform.com}}, with 48 questions in total. After five questions collecting basic demographic information, the main part of our survey consisted of two top-level sections covering, respectively, the definition and usage of bots in software development. We used the results from our interview study to devise the questions in each survey section.

\subsubsection*{Participants:}  We distributed the survey through our industry network as well as social media. Further, we invited all interviewees to participate and distribute the survey further. We received 111 complete responses, from which 59 respondents (53\%) indicated that they work as software developers, 13 (12\%) as managers,  and 39 (35\%) selected other roles (e.g., DevOps engineers, product owners, test engineers, company executives, or data scientists). Our respondents reported on average 11 years ($\pm\ 7$) of professional experience. Regarding their employing company, 61 respondents (55\%) work in cooperations (250 employees or more), 40 (36\%) in small or medium-sized enterprises (SMEs), only 6 (5\%) in startup companies, and 4 (4\%) in a single-person company. The second section of the survey was only shown if the respondent indicated that they have experience using DevBots, which 60 respondents (54\%) confirmed. Our survey took a median of 8:57 minutes to complete, and we experienced a good completion rate of 53\%.

\begin{table}
  \footnotesize
  \caption{Survey questions for the first block (bot definitions).}
  \label{tab:Definition_questions}
  \vspace{-0.3cm}
  \begin{tabularx}{\columnwidth}{lX}
    \toprule
    ID  &  Statement\\
    \midrule
    Q1 & A tool that reacts to code being pushed to a repository and triggers a build.  \\
    Q2 & Same tool, but also sends you a small report of the process as an email. \\
    Q3 & Same tool, but now it also sends you a small report of the process via chat. \\ 
    Q4 & A tool that fetches build results from elsewhere and posts a report via chat. \\ 
    Q5 & A tool triggered by a new build and runs tests on it before deployment. \\ 
    Q6 & Same tool, but now it also sends you a small report via chat. (e.g., Slack). \\
    Q7 & Same tool, but instead of starting the tests directly it prompts you via chat what actions you want to trigger. \\
    Q8 & A tool triggered by a new build and runs automated tests, the tool decides which tests to run depending on e.g. files changed or historic information. \\ 
    Q9 & A tool that creates merge requests to bugs detected in the CI pipeline, the content is based on a fixed set of known solutions and the history of accepted\slash rejected requests it has previously created for that specific project.  \\
    Q10 & A tool that creates merge requests for bugs detected in the CI pipeline. The bugfix is based on a machine learning algorithm, which trains a model using data mined from different repositories.  \\
    Q11 & A tool that in itself is a deployed entity and runs tests on another deployed entity when triggered by an engineer. \\
    Q12 & Same tool, but now it runs periodically (without explicitly being triggered). \\
    Q13 & A tool that performs static code analysis, e.g. code coverage analysis, and posts results to a separate web page. \\
    Q14 & Same tool, but now it comments the results directly to the merge request. \\ 
    Q15 & A tool that fetches the result of the analysis executed by another tool and it comments the results directly to the merge request. \\
    Q16 & A tool that fetches the result of the analysis done by another tool and send you a small report via chat.
    \\
    \multicolumn{2} {l} {A tool that helps to set up a computing environment (e.g. Docker containers) \dots}\\ 
    Q17 & \dots by reading a specification file committed to your repository containing information regarding environment, size of machines etc \\
    Q18 & \dots via a chat - you have to write the request in a specific syntax \\
    Q19 & \dots via a chat - where you can write your request in free form and the tool will merely ask for clarifications or additional information \\
    Q20 & \dots via a CLI \\
    Q21 & \dots via voice commands\\
    Q22 & A tool that reacts to merge requests to a repository, looks at the code and based on a number of parameters assigns suitable reviewers of that code. \\
    Q23 & A tool that runs a test request load on your system initiated from a CLI by an engineer \\
    Q24 & Same tool, but now it is initiated by the launch of a new version of one of the parts of your system \\
    \multicolumn{2} {X} {A tool that iterates over the repositories of your version control system to find outdated dependencies \dots } \\
    Q25 & \dots and sends you an email report when it finds one. \\
    Q26 & \dots and reports back via chat. \\
    Q27 & \dots and sends a merge request that updates this dependency. \\
    Q28 & \dots and creates a ticket for the issue in your ticket/task/bug-reporting system. \\
    Q29 & A tool that suggests improvements to code you are writing integrated into your IDE. \\
    Q30 & Same tool, but now it is integrated into your code review system. \\
    Q31 & Same tool, but now it iterates over repositories in your version control system and adds suggestions as merge requests. \\
    Q32 & A tool that while you're writing questions in a chat with someone else autonomously suggests links to other sites which it thinks contains useful information to your problem. \\
  \bottomrule
\end{tabularx}
\end{table}

\subsubsection*{Protocol:} In the first section, we briefly described variations of 11 different systems through in a total of 32 statements (see Table~\ref{tab:Definition_questions}), and asked the respondent to rate on a 5-level Likert scale whether they would consider a system with these properties a bot. Choices ranged from ``Definitely not a bot'' (1) over ``It's unclear from the description'' (3) to ``Definitely a bot'' (5). The systems were designed based on the personas and characteristics we identified in the qualitative data collection round.
For the second section of the survey, we provided 10 statements related to potential advantages of using DevBots, and asked the respondents to again rate these advantages on a Likert scale that ranged from ``Disagree'' (1) over ``Unsure'' (3) to ``Agree'' (5).

\subsubsection*{Analysis:}
To analyse the first block of questions, we mapped each described system to the three personas that emerged from interviews.
During analysis, for each respondent, we assigned points based on each answer. For example, we defined that Q4 (see Table~\ref{sec:results}) would constitute a bot for Charlie, but not for Alex and Sam. If a respondent strongly agreed with this statement (selected 5 on the Likert scale), we would add two points to the Charlie persona, but subtract 2 points for Alex and Sam. We accumulated those scores for each respondent and normalised the resulting value to the interval $[-2;2]$ by dividing it by the number of questions answered in total related to that persona. Answering ``It's unclear from the description'' or skipping a question entirely was not counted towards the total number of questions answered for this respondent. We refer to the normalized values as \textit{persona association scores}, where a score $\leq0$ would indicate no association with this persona, while higher (positive) values represent an association of increasing strength. Further, we analyse both survey sections using descriptive statistics and visually using diverging plots. For a more detailed description of the scoring procedure please refer to the replication package.

\section{Results}
\label{sec:results}

We now elaborate on the main outcomes of our study. Firstly, we introduce Alex, Sam, and Charlie
in Section~\ref{sec:resoverview}, followed by a discussion of DevBot characteristics (Section~\ref{sec:rescharacteristics}), benefits and use cases (Section~\ref{sec:resbenefits}), and challenges associated with DevBots (Section~\ref{sec:reschallenges}).

\subsection{Overview}
\label{sec:resoverview}

A key goal of our study was to systematically identify and categorise what qualities, characteristics, or properties turn a ``Plain Old Development Tool'' (PODT) into a ``bot'' in the eyes of practitioners.
 We find that there are fundamentally three different groups among our interviewees, depending on how they define DevBots for themselves. We name and identify three personas, i.e., practitioner archetypes with different expectations and motivations~\cite{pruitt:03}.

First, the \textbf{chat bot persona} (\underline{C}harlie) primarily equates bots to tools that communicate with the developer through a natural-language interface (typically voice or chat), while caring little about what tasks the bot is used for or how it actually implements these tasks. Virtually any PODT can become a DevBot for Charlie if it exposes a natural-language interface. Unlike the other personas, for Charlie the ``bot'' is normally only the interface to an existing system, not the system itself. A typical example of DevBots for Charlie are bots for ChatOps, i.e., bots that execute team or environment management tasks based on commands written in a platform such as Slack. One interviewee strongly associated with the Charlie persona is I5:

\interviewquote{The fact that I chat with it makes me think more of it as a bot.}{I5}

Second, the \textbf{autonomous bot persona} (\underline{A}lex) defines bots primarily as tools that work on their own (without requiring much input from a developer) on a task that would normally be done by a human. Hence, not every scheduled script execution is a DevBot for Alex, but a script executing a development task such as closing bugs or welcoming newcomers to a project may be. Alex places particular emphasis on the highly independent nature of a bot --- once configured for a project, a DevBot can sense autonomously when its service is needed (e.g., by monitoring open bugs in an issue tracker). Many widely used GitHub bots, such as Dependabot\footnote{https://dependabot.com}, can be seen as illustrative examples for how Alex thinks about DevBots. An example from our interviewee population is I4:

\interviewquote{The trigger shouldn't be a human taking initiative. It should trigger on something, something else than someone telling it to do so.}{I4}

Third, the \textbf{smart bot persona} (\underline{S}am) distinguishes DevBots from  PODTs primarily through how ``smart'' (technically sophisticated) a tool is. Sam cares less about how the tool communicates, but more about whether it is unusually good or adaptive at executing a task.
Sam's view of DevBots is in some ways similar to Alex'. However, while a relatively simple autonomous tool may be a DevBot for Alex, Sam expects more from a bot.
For Sam, a PODT may, for instance, simply do predefined text replacements in a code base, while a DevBot understands the code syntax and is able to execute higher-level tasks (such as fixing bugs) even on code that does not exactly follow a rigid predefined pattern. For this persona, bots are often strongly associated with machine learning and/or advanced program analysis techniques. I21 is an example interviewee:

\interviewquote{They're sort of these agents that have triggered on some events that you may not even be aware of and they perform an interpretation of the world that you don't know the rules of, and then they perform an action that you may notice.}{I21}

\begin{figure}[h!]
  \centering
  \includegraphics[width=\columnwidth]{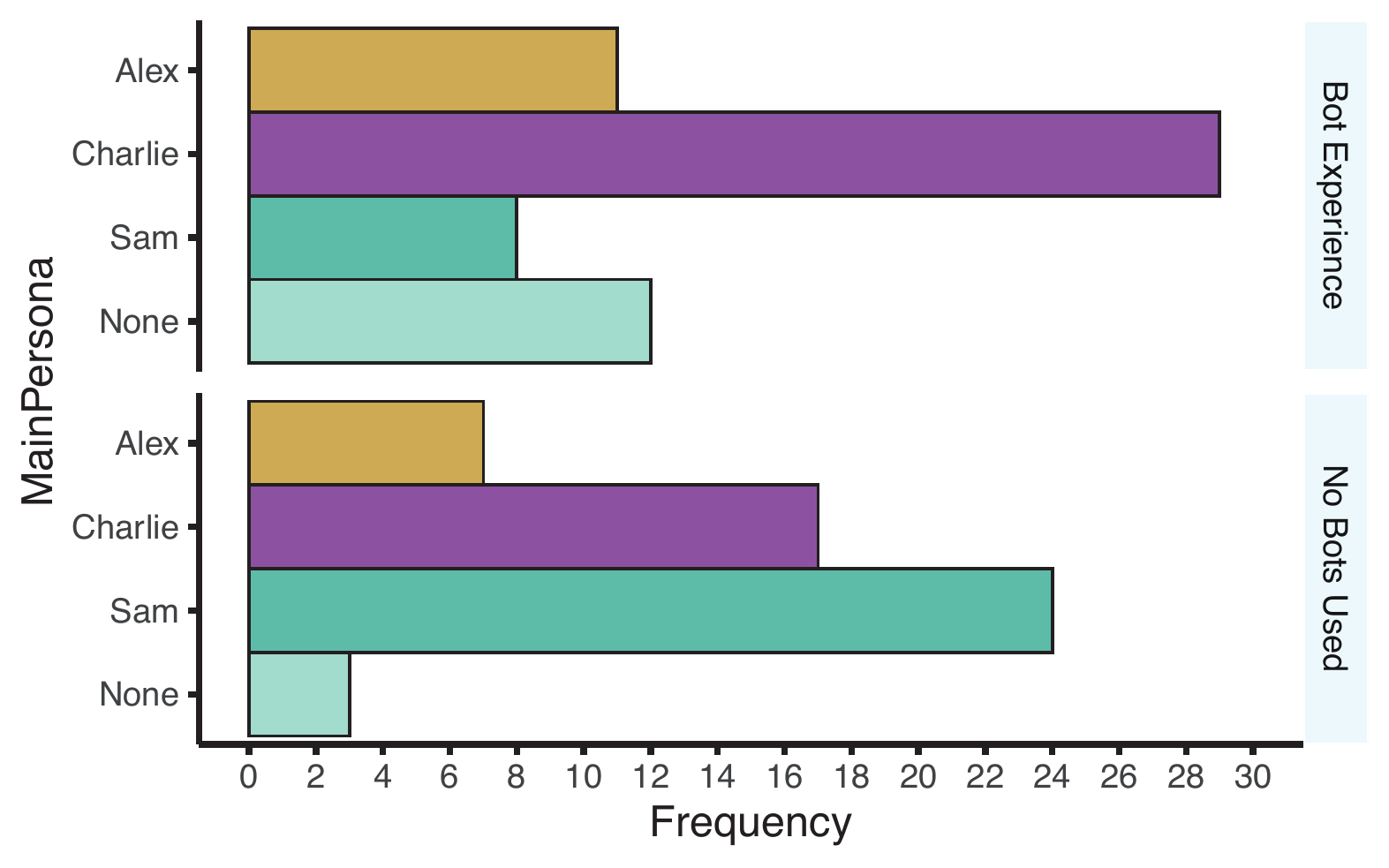}
  \caption{Frequency of survey respondents predominantly associated to Alex, Charlie, and Sam.}
  \label{fig:frequency}
\end{figure}

Our survey data allows us to estimate how many respondents are associated with each persona (see also Section~\ref{sec:method}). Figure~\ref{fig:frequency} depicts the frequency of each persona in our survey data, as defined through the highest persona association score obtained for each respondent. Two respondents had tied values for highest persona scores. We then randomly assigned the corresponding respondent to one of their tied personas. We classify respondents as belonging to no persona (``None'') if their score is negative for all personas. Moreover, we distinguish between respondents that reported experience with using DevBots and respondents that do not.


In total, we classified 18 (16\%) respondents as predominantly Alex, 32 (29\%) as Sam, 46 (41\%) as Charlie and 15 (14\%) respondents are not associated to any persona. However, there are evident differences between respondents that are (or were in the past) actively using DevBots and those that do not. Among actual bot users, Charlies are a clear majority (29 of 60, or 48\%) followed by Alex (11, or 19\%) and Sam (8, or 13\%). It is also interesting to observe that among actual bot users, we found substantially more respondents not clearly associated to any persona.
One possible reason for this phenomenon may be that most existing, industrial-strength DevBots are chat bots or simple automation tools. There are few, if any, off-the-shelf systems available that a Sam \emph{could} easily use --- our interviewees associated with Sam exclusively work for larger corporations, which have invested considerable resources to develop smart bots custom for their own development environments.

\begin{figure}[h!]
  \centering
  \includegraphics[width=\columnwidth]{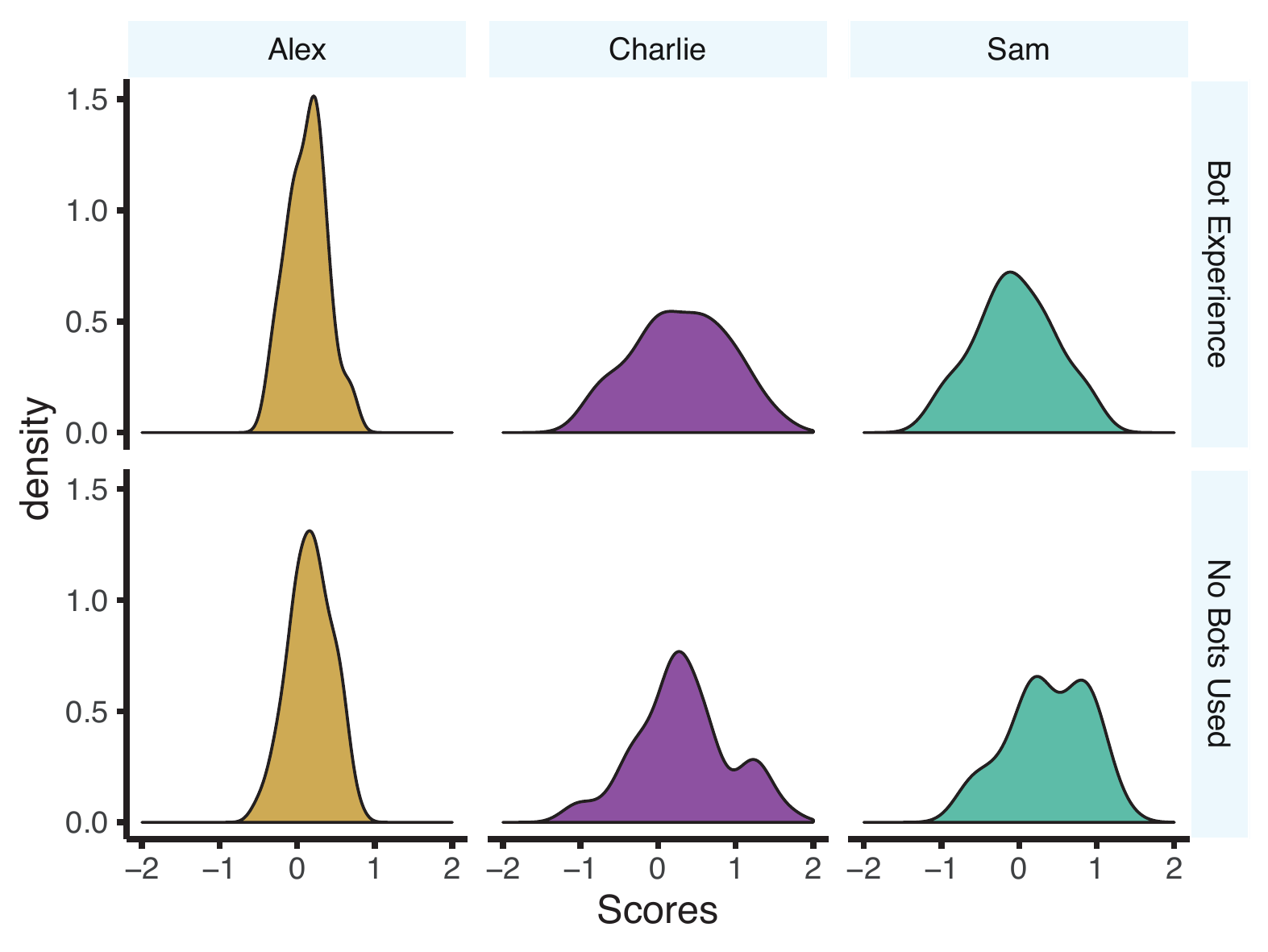}
  \caption{Density diagram of persona association scores for Alex, Charlie, and Sam.}
  \label{fig:density}
\end{figure}

However, this high-level result masks that personas are not rigid either-or categories, and many respondents (as well as many interviewees) combine characteristics of multiple personas. Indeed, we find that many respondents have association values $>0.5$ for two separate personas. In Figure~\ref{fig:density},
we show a density diagram of the persona association scores, again split by whether
respondents reported to actually use any DevBots.

We observe that few respondents map rigidly into a specific persona. Instead, most combine traits of multiple personas, with association scores largely distributed between -1 and 1 (with a slight trend towards positive values, which may be due to self-selection bias of our study population). Scores in the Alex persona are more neutral than for the other personas (the majority of respondents has scores between -0.5 and 0.5, indicating no particularly strong opinion). This may be because characteristics associated with Alex are, by and large, rather uncontroversial. We again observe noticeable differences between respondents that are actually using bots and those that do not: bot users have higher Charlie and lower Sam scores, and, counter-intuitively, more negative scores indicating no strong association to any persona.

%
%

\subsection{Characteristics of DevBots}
\label{sec:rescharacteristics}

\begin{figure}[h!]
  \centering
  \includegraphics[width=0.7\columnwidth]{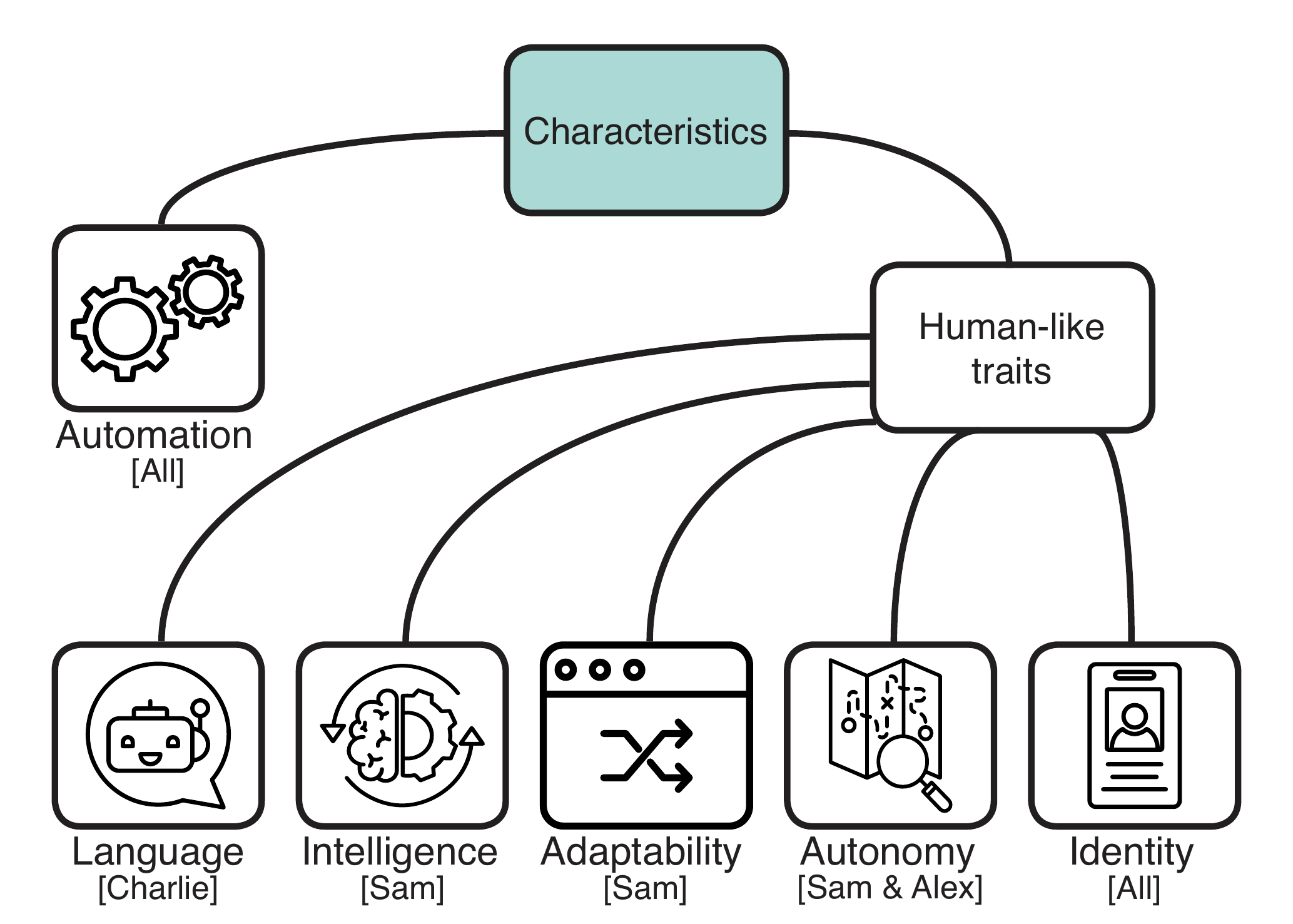}
  \caption{Characteristics of DevBots.}
  \label{fig:characteristics}
\end{figure}

Figure~\ref{fig:characteristics} shows an overview of characteristics emerging from our interviews. For all three personas, a DevBot is a tool that exhibits some \emph{human-like traits} and which \emph{automatically} executes a task.

What kind of tasks DevBots are used for varies between fairly complex and technically involved operations (e.g., setting up test environments, doing version updates) and simply fetching information from one tool and displaying it in another. However, ultimately, all interviewees (and all personas) agree that fundamentally a DevBot carries out tasks that would otherwise be done by a human. As I2 puts it:

\interviewquote{It's sprung from something that we did manually previously and [now] we have a bot do something that we could do ourselves.}{I2}

\subsubsection*{Identity [all]}
Charlie, Alex, and Sam all see different human-like traits as the main characteristics of DevBots. However, they all agree that a quasi-human identity, e.g., a human-sounding name or cute profile picture, are a common characteristic that distinguishes DevBots from PODTs. Research has shown experimentally that bots with a human identity can evoke emotions in human users, even if they are aware that they are interacting with a bot~\cite{lee:19,Nass2000HCI,KIM2012241}. Some interviewees have expressed that the bot identity is sometimes designed explicitly to evoke specific emotions or steer how the bot is perceived by developers:

\interviewquote{(..) you put an icon on it and some googly eyes (..) doing that actually changes the dynamic of how people react to it.}{I21}

\subsubsection*{Autonomy, Intelligence, and Adaptability [Alex and Sam]}
For Alex, the central human-like trait of a DevBot is its autonomy. In practice this autonomy can range from very simple (e.g., a bot that runs at predefined intervals) to a complex system of triggers based on monitoring data or external state. For instance, I4 indicates that an automation-focused tool in addition needs to also exhibit some autonomy in its decision-making to become a DevBot:

\interviewquote{If I have like a Python script that checks my service and I run it once I wouldn't say that's a bot.}{I4}

For Sam, a combination of three different characteristics (intelligence, adaptability, and autonomy) appears to be crucial for a tool to be perceived as a DevBot. In terms of adaptability, some interviewees have stated that what really distinguishes a DevBot from a PODT is the ability to learn and improve, without explicit reconfiguration or retraining by a human:

\interviewquote{They can also learn something from the interaction and then kind of improve themselves in a way.}{I17}

Different DevBots may adapt in different ways --- they may improve in their task, better customise their service to specifics of the project, or learn the preferences of different human developers. However, this characteristic appears to be underexplored at the time of study. Most bots currently in use do not actually implement any explicit feedback loops that would allow them to improve on their own.

In terms of intelligence, our interviewees often struggled to express their views clearly, as all participants were acutely aware that human-level intelligence is out of reach of any current bot. However, responses indicated that interviewees thinking like Sam expect a degree of smartness, context-awareness, or fitness for a task from a DevBot that goes beyond normal tools (or typical Alex-style bots). This focus on intelligence naturally links the idea of DevBots to advances in artificial intelligence. As I3 puts it:


\interviewquote{I realise that there is no intelligence, there's rules. It's just that some systems have more rules than others.}{I3}

%
%
\subsubsection*{Language [Charlie]}
For Charlie, an ability to produce or mimic understanding of spoken or typed natural language is key. Virtually any system can be perceived as a bot as long as it interfaces with human developers via a natural language interface:

\interviewquote{(..) if there is some kind of automated response in natural language (..), regardless of if it's bidirectional, I think that qualifies as a bot for me.}{I13}

Such bots normally integrate with existing communication tools (e.g., Slack), and much of their perceived value lies in this tight integration with communication systems that human developers also use for status updates and synchronisation on work tasks. However, how exactly a tool surfaces information is not important for Charlie, as long as it does so via natural language.



%

\subsubsection*{Survey Results}
\begin{figure}[h!]
 \centering
 \includegraphics[width=\columnwidth]{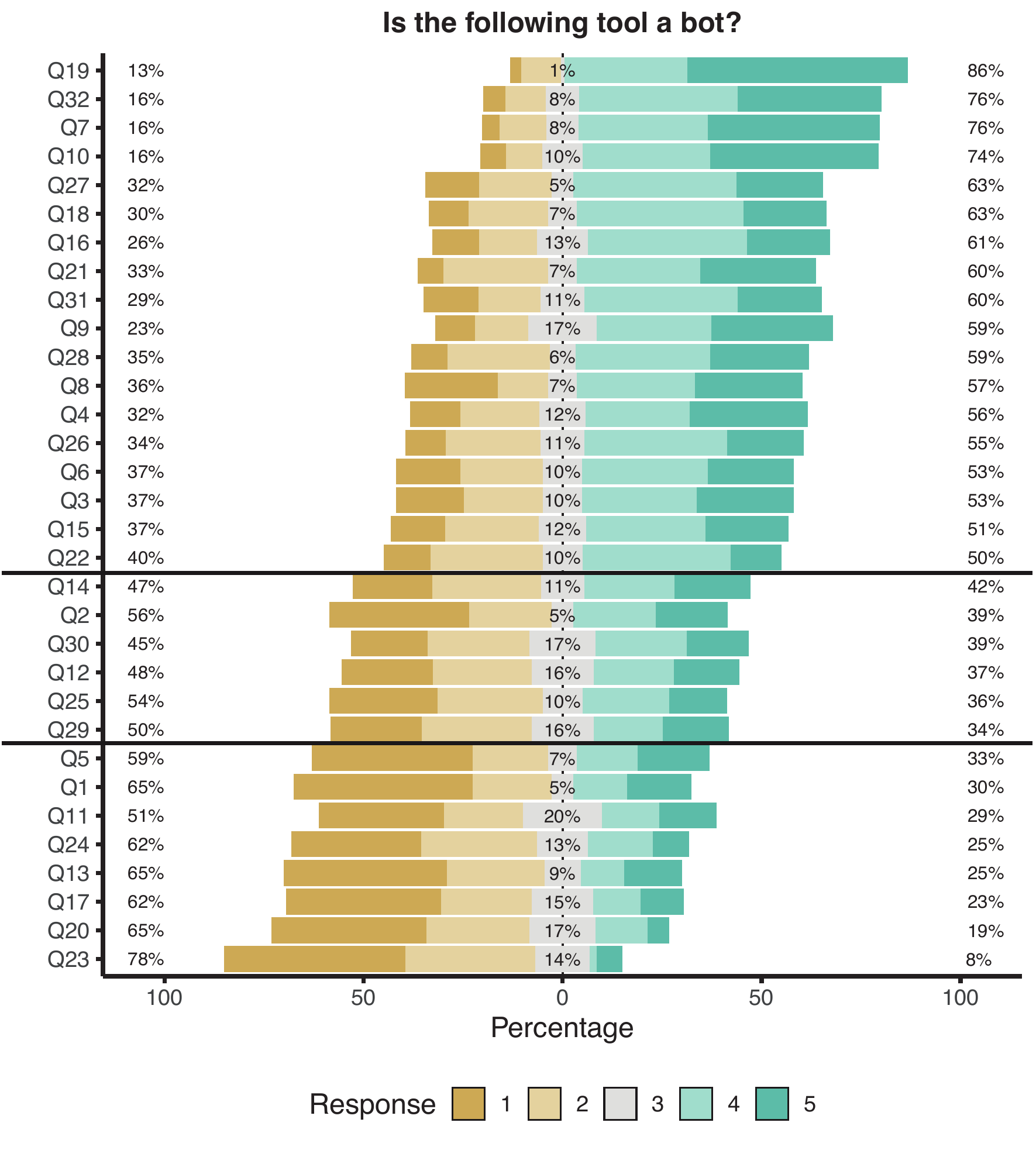}
 \caption{Diverging plot for all 32 questions related to bot definitions. Refer to Table~\ref{tab:Definition_questions} for the mapping of identifiers to actual questions.} 
 \label{fig:sur_botdefinition}
\end{figure}

We now revisit which systems our respondents predominantly considered bots (see also Table~\ref{tab:Definition_questions}). Figure~\ref{fig:sur_botdefinition} shows a diverging plot of all 32 questions, ordered by agreeance across all respondents.

Eight of the systems were specifically designed \emph{not} to describe bots, as they lack all of the characteristics that emerged from the interviews. Indeed, these were the lowest ranked, all ending up with $>50\%$ of the participants selecting one of the negative options (1 or 2 in our Likert scale). Next comes a group of systems where $<56\%$ of respondents selected one of the negative options, but $<50\%$ selected one of the positive options. These systems include several Alex- and Sam-style DevBots, which can be interpreted either as a DevBot or as a part of existing standard tools (e.g., code review, CI/CD pipeline, or IDE). Finally, in the top section we find the group of tools where more than $50\%$ selected one of the positive values (4 or 5). Here we find all the tools communicating via chat, no matter how one directional the conversation is. We also find the Sam- and Alex-style DevBots that produce non-trivial code snippets or analysis. Interestingly, we observe that the same tool may be perceived more bot-like if it is standalone than if it is integrated into an existing system (e.g., compare Q29 and Q31). Another interesting takeaway is that a tool that merely integrates result from other tools in a chat or as comments on merge requests is considered more bot-like than if the producing tool itself communicates.(see for instance Q3 versus Q4, or Q14 versus Q15).

However, in general it is important to keep in mind that the personas of our respondents influences this analysis, and 43\% of total respondents were classified as predominantly associated to Charlie. Hence, it is unsurprising that systems that mention communication via chat are ranked high in Figure~\ref{fig:sur_botdefinition} (e.g., Q19, Q7, Q32).

\subsection{Potential Benefits and Bot Use Cases}
\label{sec:resbenefits}

\begin{figure}[h!]
  \centering
  \includegraphics[width=0.7\columnwidth]{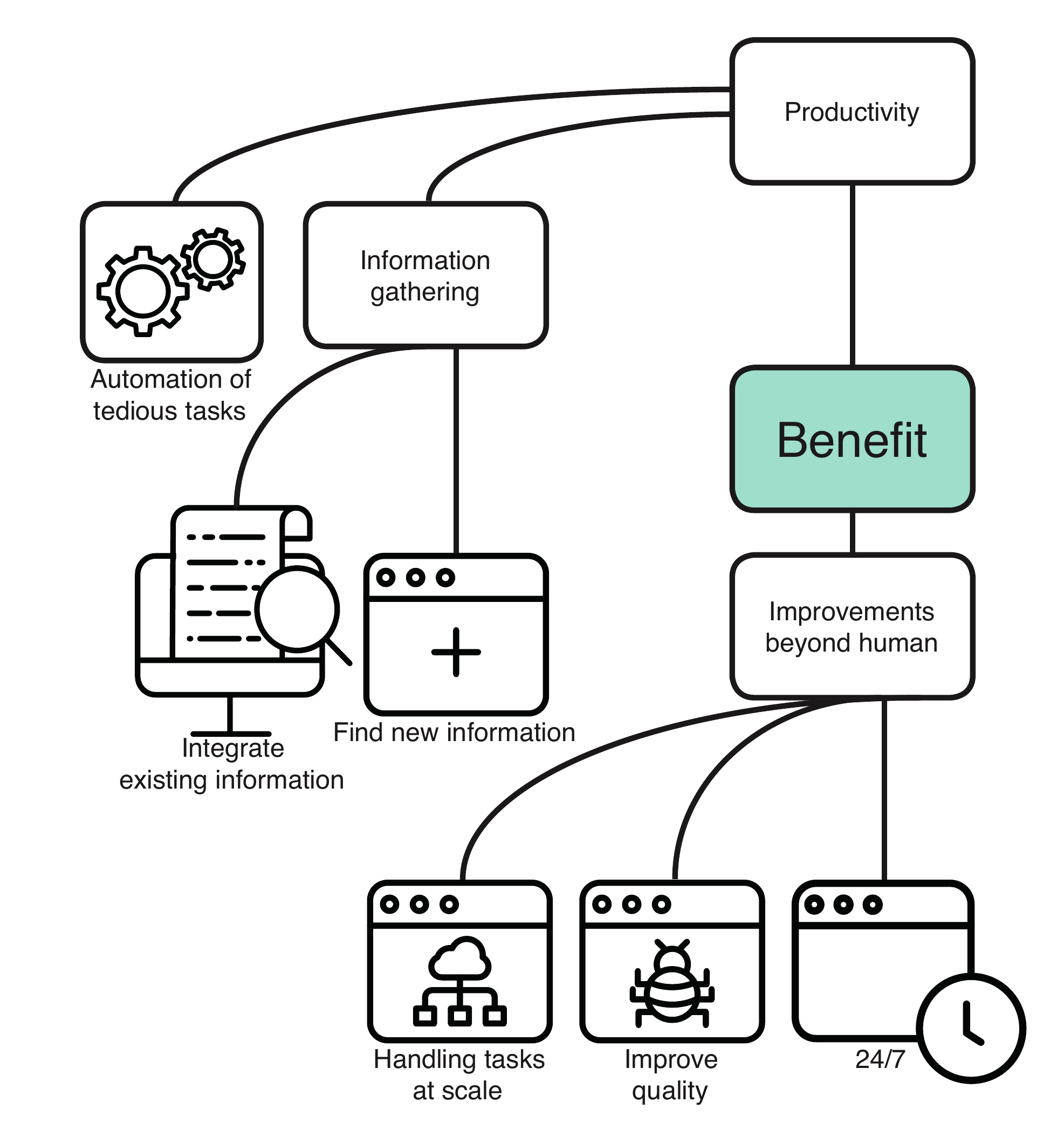}
  \caption{Potential benefits of DevBots}
  \label{fig:why}
\end{figure}

As discussed in Section~\ref{sec:rescharacteristics}, DevBots generally work on tasks that would otherwise have to be done by humans. This can be for one of two reasons: either to improve productivity (all personas), or because using the bot in some way improves the quality of the work or enables use cases for which humans are not realistically suitable (Alex and Sam). An overview of types of potential DevBot benefits is given in Figure~\ref{fig:why}.

\subsubsection*{Productivity [all]}
In terms of productivity, DevBot usage is frequently motivated by a need for developers to spend less time on activities that could be automated. While all personas agree that improving productivity in an abstract sense is a main driver to adopt DevBots, how a bot helps with productivity is seen differently. In our interviews, this emerged as two main branches: automation of tedious tasks and information gathering.

For Alex, the productivity gains lie mostly in the automation of simple, but time-consuming, annoying, or otherwise tedious tasks. However, even for Alex, DevBots may sometimes help to integrate existing information, hence alleviating a developer from having to collect the data themselves:

\interviewquote{So before that (..) we could see manually if the coverage is dropped. (..) this cool bot is able to integrate well into the process. So I like that I don't have to go to a different websites}{I13}

For Charlie, information gathering (both integrating existing information or generating information that would otherwise be hard to acquire) is the main productivity gain associated with DevBots. They collect information from heterogeneous systems under a common umbrella (often a team chat), where it can be acted upon instead of having to log to the original systems. For example, in ChatOps, an on-call engineer can be more productive when all required troubleshooting information is available directly through a bot. Notably, for I18 this has the advantage of circumventing company firewalls, allowing them to act quicker when out of office:

\interviewquote{If something happened, if I am for example in the cinema or on the road, I don't have my VPN ready so I can't do the query myself or see the dashboard. Everything is locked behind the firewalls, let's say. And I can use such a bot to get a screen shot, get some data out of there so I can triage the incident faster.}{I18}

Similar to Alex, Charlie may also use DevBots for the automation of tedious tasks, such as setting up build environments, and other scaffolding. For example, I8 has a DevBot for managing their production infrastructure.
%




Developers thinking like Sam also care about productivity. However, for them, DevBots are less about automating tedious tasks or simple information integration (this is the realm of PODTs for Sam). Instead, they improve productivity by looking into the vast amount of data modern systems generates and gaining new insights or finding patterns that humans might overlook. As I21 puts it:

\interviewquote{(..) discovery of information that is there, but humans may not notice or may have a harder time getting to it.}{I21}

\subsubsection*{Improvements Beyond Human [Alex and Sam]}
For Sam and Alex, there are often other benefits orthogonal to productivity. For them, DevBots may actually do some tasks better than humans. Three classes of such advantages emerged from our interviews. For Alex, this includes handling tasks 24\slash 7 (i.e., at any time, and without requiring lead time) as well as handling tasks at scale (i.e., acting upon a high volume of tasks quickly). For both Sam and Alex, a third class of benefits is improving quality.
Using DevBots may increase consistency and eliminate human error, as bots can reliably solve the same tasks in the same manner, without being subject to differences in coding style, taste, or developer qualification. In our study this surfaced in various ways, from relieving developers from having to care about code formatting (I19) to skipping training on company conventions (I21):

\interviewquote{You know at some point, not everybody would be exactly familiar with all the ways to do work. So you want to introduce some level of consistency so that people don't have to think about it and get through training every three months.}{I21}

More broadly, we have observed that, for Alex and Charlie, DevBots are often supporting tools -- they are useful, but not necessarily crucial for the development process. For Sam, on the other hand, the gains from using DevBots can be dramatic, with bot adoption transforming how software is being built:

\interviewquote{I think our ecosystem has evolved to the point where they are pretty integral.}{I7}

\subsubsection*{Survey Results}
\begin{figure}[h!]
  \centering
  \includegraphics[width=\columnwidth]{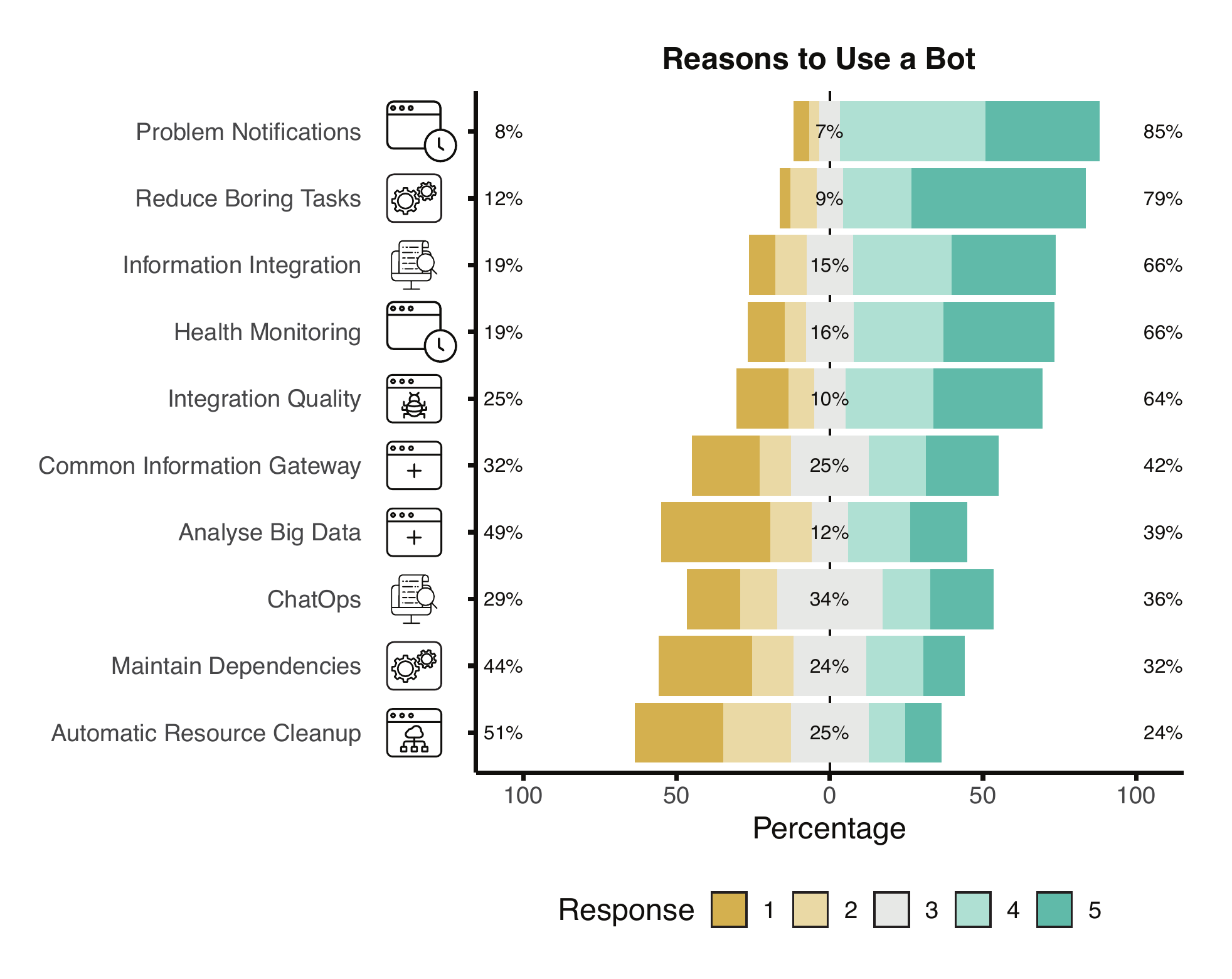}
  \caption{Diverging plot of the perceived importance of different DevBot use cases.}
  \label{fig:sur_botusage}
\end{figure}

In the survey, we explicitly asked how important the respondents rated ten reasons for DevBot usage that emerged from our interviews. Figure~\ref{fig:sur_botusage} depicts the answers as a diverging plot. The icon indicates what type of benefit (see Figure ~\ref{fig:why}) the specific statement corresponds to. Note that this question was only presented to respondents who indicated that they are, or have been, actively using DevBots in the past.

85\% of respondents agreed or strongly agreed that a reason to use DevBots is to be quickly aware of problems (``Problem Notifications''), followed by a desire to automate tedious tasks (``Reduce Boring Tasks'', 79\%). In turn, 66\% agree to use DevBots to monitor the ``up status'' of services (``Health Monitoring'') and to have all information in one place (``Information Integration''), whereas 64\% are interested in monitoring, specifically, code integrations (``Integration quality''). For the remaining five reasons, less than half the survey population has agreed that they are a reason to use DevBots. These are using DevBots as a common gateway for all system information (``Common Information Gateway'', 42\%), to collect and analyse large chunks of data (``Analyse Big Data'', 39\%), ChatOps (36\%), to maintain their software dependencies (32\%), and to automatically clean up system resources, such as unused cloud instances (``Automatic Resource Cleanup'', 24\%).

\subsection{Challenges}
\label{sec:reschallenges}

Given that Sam, Alex, and Charlie have fairly different views on what a DevBot is, their challenges in DevBot usage also differ. We group the challenges that emerged from our interviews in four groups: interruption and noise, trust, usability, and others.

\subsubsection*{Interruption and Noise [all]}
A potential problem pertaining DevBots for all personas is the trade-off between timely notifications or requests for feedback on the one hand, and frequent interruptions (or simply producing too much information) on the other.

For Charlie (who often uses bots as means of information integration), this often manifests as DevBots creating noise in human communication channels (e.g., overloading Slack channels with information), or just posting so frequently that human developers stop paying attention. This can be alleviated through a bot design or configuration that is mindful of whom to notify of which events:

\interviewquote{Because the Slack bot, for example, is just posting messages into a general group where many of us are subscribers. So everyone gets all the notifications, and probably most of us are only interested in a few.}{I14}

For Alex, noise manifests similarly, for instance through GitHub bots posting too many messages in code review systems. Oftentimes, the problem is not one particular bot that is too verbose, but rather that too many bots are simultaneously active and posting status updates.

However, for Sam, DevBots often interact more proactively and directly with the developer. Such systems need to carefully evaluate how often, and when, they should interrupt the developer with suggestions or requests for further input. A good bot waits until a developer is ready for feedback. I7 has indicated that, for this reason, a bot that interfaces with the code review system is preferrable to one that integrates into the IDE:

\interviewquote{Once I get to a point where I'm doing code review (..) at that point, I'm much more happy to have a proactive tool come and edit my code for me.}{I7}

\subsubsection*{Trust [Alex and Sam]}
For both Alex and Sam, trusting the DevBot to act appropriately can be an issue. Most interviewees are not troubled trusting a bot to correctly execute very simple and well-defined tasks, such as restarting a server or re-running a test. However, Alex typically is sceptical regarding DevBots that actively modify code or execute other operations that require significant understanding of the system or environment.

For Sam, trust is a complex issue with multiple trade-offs. On the one hand, the very definition of what a DevBot is for Sam requires bots to come up with unanticipated solutions to non-trivial problems, and implement them with wide-ranging autonomy. On the other hand, Sam is also aware that even small problems can cause developers to lose trust in their DevBots. A key solution to these issues is the existence of a reliable test infrastructure. As I6 puts it, if you have good tests, there is no reason not to allow a DevBot to, for instance, modify source code:

\interviewquote{I think if you start using bots to automate tasks in your project it's very dependent that you have a lot of tests. So if the projects are fully tested (..), I would trust them to do most things}{I6}

However, even if problems can be detected timely, DevBots producing too many broken builds, failed deployments, or irrelevant warnings will fall out of use quickly. I7 emphasises the need for DevBots to be aware of their own limitations, and only act if they are sure that the proposed action is correct. Otherwise, it is ultimately better to alert the developers of the problem than to enact a solution that may turn out to be broken:


\interviewquote{(..) we don't have false positives, but we can put up with some false negatives because we can identify those cases and fix them by hand.}{I7}

This is true even for our interviewees using DevBots for infrastructure monitoring. Bots producing an excessive amount of false positives (i.e., monitoring alerts that do not relate to actual problems) are perceived as less useful than more conservative ones, even at the cost of occassional false negatives (missed alerts).

Another approach that emerged from our interviews is to conduct a risk assessment, and to set a threshold for how high-risk tasks an automated tool is allowed to handle:

\interviewquote{(..) model some kind of impact, severity or risk of operation that the software can evaluate itself (..) if I can set some thresholds on some critical conditions, then I'm fine with it doing [tasks] without interaction.}{I12}


Interestingly, a side effect of environments with mature and trustworthy DevBots appears to be that bots can become an implicit authority in the tasks that they handle. I7 has reported that one challenge for their team is that human developers are not sufficiently questioning the actions of DevBots anymore, but simply assume that the bot's way of solving an issue is what's best. This is often unwanted -- if an automated process is reliable enough that it requires no human oversight, it would be implemented directly in the infrastructure or toolchain. If automation is implemented as a DevBot, the assumption is that there are corner cases where the DevBot solution will break or be suboptimal, and which require the oversight of a human developer:


 \interviewquote{The reason we put them in bots [is] because you want people to think about the result and decide whether they should apply or not. It's been interesting to me to observe how often people don't actually do that. They just assume what the bot says must be correct and move on. }{I7}

%

\subsubsection*{Usability [Charlie]}
Charlie wants to interact with bots through a natural language interface. However, the more sophisticated the language parsing of the bot becomes, the more conversations are perceived as ``natural'', but the less obvious it is to see what actions will be the outcome of any particular conversation, leading to usability problems:

\interviewquote{It was very difficult for you from the outside to understand why it showed one particular matching instead of another.}{I5}

One simple fix for this problem that is already widely adopted by bot makers is to have the DevBot ask for confirmation before starting to work on a task:

\interviewquote{I would often want the bot to repeat back to me what it has understood in terms of instructions, given that there is risk for misunderstanding.}{I5}

Some bots alleviate this problem by relying on a simplistic natural language syntax, and require typing or voicing exact sentences rather than parsing intents. However, in this case, the developer has to remember what exactly to say or type to trigger which functionality, not unlike remembering shell commands. Such bots are perceived as less powerful, less useful, or are sometimes not considered bots at all.

%

%

\subsubsection*{Other Challenges}
Other challenges when using DevBots are more specific to individual types of bots or usage scenarios.
I10 sees a challenge in explaining to non-technical stakeholders what a bot actually does. Some interviewees which operate a large number of DevBots see the problem that bots start interfering with each other's goals. For instance, I21 reported an example where the team used a bot that opened work items for upcoming tasks, and another bot that quickly closed the same work items because it was seeing too little activity on them.

\interviewquote{So there is a problem of having many bots: they interact with each other.}{I21}


Finally, it is interesting to observe that \emph{none} of our interviewees has expressed fear to be replaced or made redundant because of current or future DevBots. So far, developers struggle to imagine a world where software development is not predominantly a human activity, in which bots -- even smart Charlie-style bots -- only play a supporting role.

\section{Discussion}
We now elaborate on the broader implications and lessons learned of our research for practitioners and software engineering researchers.\\

\begin{figure}[h!]
  \centering
  \includegraphics[width=\columnwidth]{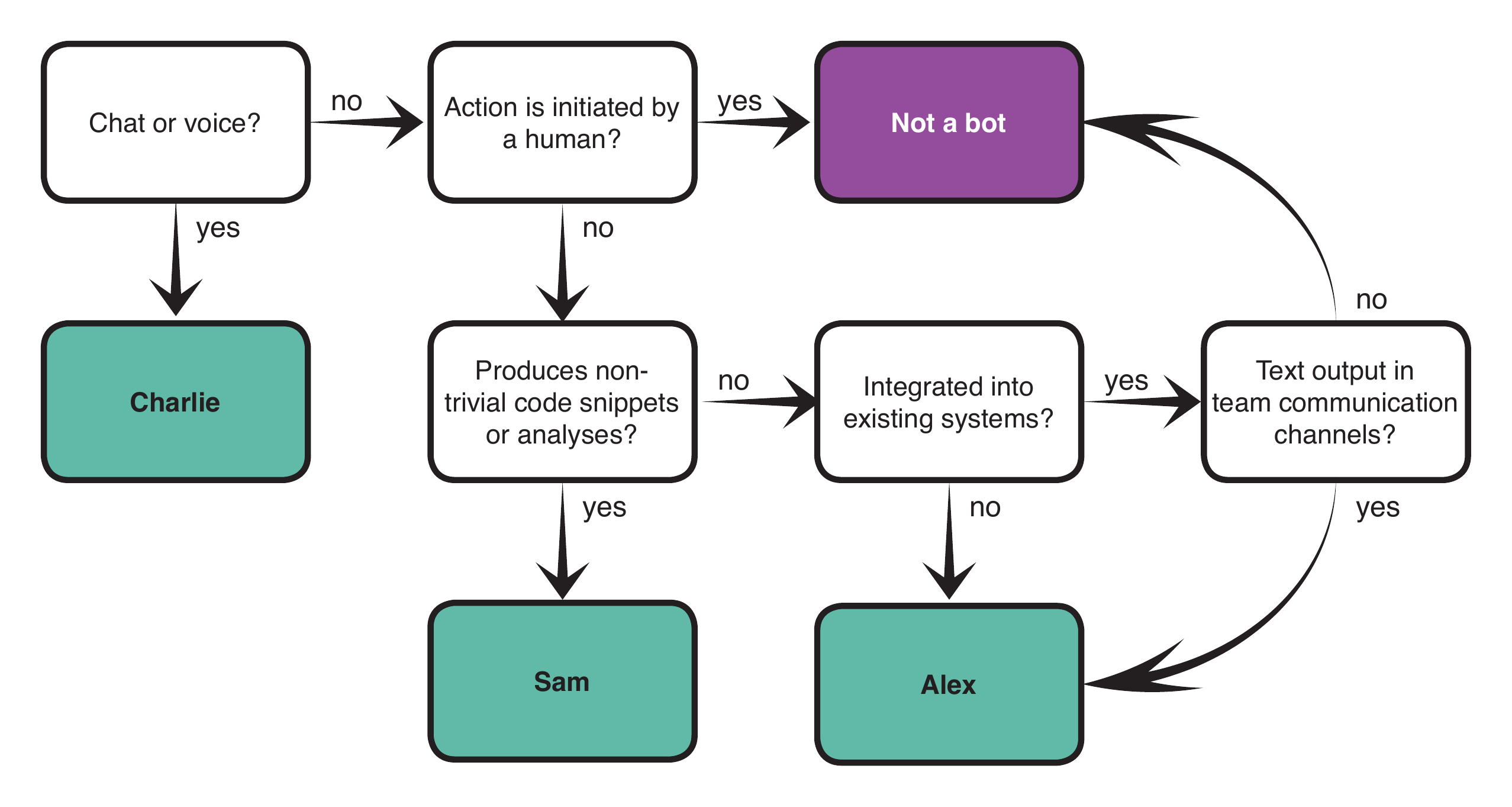}
  \caption{Simplified model to classify DevBots.}
  \label{fig:flow_chart}
\end{figure}

\textbf{What is a DevBot, really?}
The core question our research attempts to answer is what a DevBot actually is.
This question is crucial for the budding field of bot-based software engineering, as a meaningful scientific discussion relies on a shared definition of the subject of study. Figure~\ref{fig:flow_chart} depicts, in slightly simplified fashion, a flow diagram that has emerged from our interviews. This model can be used to decide on a high-level whether a given tool is likely to be considered a bot, and for which persona.

\textbf{Implications for Researchers.}
Future empirical work on bot-based software engineering can use our results to contextualise what \emph{types} of DevBots their work actually targets. This would improve clarity and manage expectations. For example, within our framework, the study of Wessel et al.~\cite{wessel:18} reports on the prevalence of Alex-style bots in open source systems, making clear that their main focus is on relatively simple autonomous tools (rather than chat bots or machine learning based tools). Although not the core of this work, our results can also be used to provide more objective inclusion criteria for future empirical studies, i.e., our results make it easier to objectively answer the question whether a given tool should be considered a bot in the context of a given study.

Our results related to perceived benefits and challenges can also be used to direct future research that aims at the development of new types of DevBots (e.g., future work similar to Matthies et al.~\cite{matthies:19}) to the areas that will actually benefit developers most, as well as help with the evaluation of new DevBots. For example, our results show that a very small amount of false positives is crucial to the practical adoption of Sam-style DevBots, indicating that this is an area that future studies should be focusing on.

\textbf{Implications for Practitioners and Bot Builders.}
When conducting interviews, we have observed that confusion related to what really defines DevBots is not unique to the academic community. We have seen the same questions raised by practitioners, and multiple participants have expressed that our interviews have challenged their own understanding of the domain. Hence, our framework can also be useful to practitioners.

Further, we have observed that both Charlie and Alex-style DevBots are already widely available as off-the-shelf systems, allowing developers to quickly adopt, e.g., mature ChatOps or CI bots. However, there appear to be very few (or none) general-purpose Sam-style bots available to choose from. Our interviewees associated with Sam exclusively worked for large software-intensive corporations, which invested substantial development effort into building and maintaining bespoke bot infrastructures. This is also consistent with the results presented in Figure~\ref{fig:frequency}, which showed that most survey respondents associated with the Sam persona are not currently using any bots.

However, we have also seen that the gains from adopting Sam-style DevBots can be dramatic: our interviewees have reported that adopting these bots has transformative effects that go beyond simply automating tedious tasks. Hence, we see an opportunity for smart bots that are not custom-built for a specific project or corporation, to bring at least some of these benefits to the wider developer community. However, our results also show that these tools need to be carefully designed to see practical adoption (e.g., related to trust, usability, or at which point in the development process they surface their work).

\section{Threats to Validity}

%


Designing our research as a mixed-method study allowed us to triangulate the results obtained through interviews with quantitative survey data. However, a number of limitations remain. In terms of \emph{external validity}, we cannot claim that our study population (for both methods) is representative of software engineers in general, as both populations have been sampled through our personal network (convenience sampling). To mitigate this threat, we selected interview participants to cover companies of different sizes and in different domains. However, given our sampling method, a majority of interviewees are working in the same broad geographical region. For the survey, we did not collect detailed company or geographical information to prevent de-anonymising some participants. However, we have to again assume that the respondent population is relatively homogeneous in terms of geographical distribution. Further, a voluntary survey design is always susceptible to self-selection bias: respondents uninterested in using bots for software development are unlikely to participate in our study. This may also explain why we have received relatively few responses not strongly associated to any of our personas (see Section~\ref{sec:resoverview}).

In terms of \emph{internal validity}, a threat is that we were, through our previous interest in the field, pre-exposed to existing research (e.g., existing bot taxonomies~\cite{erlenhov2019botse,lebeuf2018ieee,paikari:18}), which may have biased our interview design. Hence, as \emph{contruction validity} threat, we may have missed potential characteristics (or, worse, entire personas) because they have not been featured prominently in previous research. However, given that most of our survey respondents mapped well into the three personas identified in this work, we judge this threat to be low. Nonetheless, given our study design, we cannot claim that the identified list of personas, characteristics, benefits, and challenges is necessarily complete. Finally, we have observed that participants tended to become more conservative in their ratings while progressing through the questionnaire. This threat could have been mitigated by randomising the order of question blocks. Unfortunately, our survey tool did not support this feature.



\section{Conclusions}
This paper investigates the characteristics, benefits, and challenges that define a bot in software development (DevBots).
Our analysis shows that no single definition fits all practitioners. Instead, we identify three different personas, each associating distinct characteristics with the term ``bot''. The chat bot persona (Charlie) mostly sees DevBots as information integration tools with a natural language interface, while for the autonomous bot persona (Alex) a DevBot is a tool that autonomously handles repetitive tasks. Lastly, for the smart bot persona (Sam), the defining feature of bots is its degree of ``smartness''.
The personas also associate different benefits and challenges with DevBot usage. Notably, Charlie uses bots primarily to have access to all information in one place, but struggles with usability. For Alex and Sam, DevBots are used to improve productivity by automating tasks (simple in the case of Alex, more sophisticated for Sam), but trusting those bots to trigger correct actions can be a challenge. In turn, all personas, to some extent, struggle with interruptions or noise produced by bots.
Finally, we have observed that there are currently few, if any, general-purpose Sam-style DevBots available. We consider this problematic, as these more sophisticated DevBots also promise higher gains than the currently prevalent Alex- or Charlie-style bots.
Our contributions include a framework to define and describe existing and future DevBots. This framework  supports future research and applications in the field by allowing researchers or practitioners to target specific types of bots or meet the persona's corresponding expectation.
As future research, we aim to further investigate the boundaries and relationships between personas in order to identify delineating factors that can refine our framework and reveal insights about creating and adopting bots in software engineering.

\section*{Acknowledgements}
This research has been funded by Chalmers University of Technology Foundation and the Swedish Research Council (VR) under grant number 2018-04127 (Developer-Targeted Performance Engineering for Immersed Release and Software Engineers).
Icons used in figures made by Icongeek26 from www.flaticon.com
\bibliographystyle{acm}
\bibliography{bibliography}

\end{document}